\documentclass{article}
\usepackage{hiph-preprint}
\usepackage{graphicx}
\volnumber{22} \issuenumber{1} \edyear{2005}                             %|
\frompage{000} \topage{000}                                              %|
\recrevdate{1 January 2005}                                              %|
\def\rightmark{High $p_{T}$ $\gamma$-hadron and $\pi^{0}$-hadron correlations in $\mathrm{Au}+\mathrm{Au}$ collisions}
\def\leftmark{J. Jin for the PHENIX Collaboration}

\title{High $p_{T}$ $\gamma$-hadron and $\pi^{0}$-hadron correlations in \\
$\sqrt{s_{NN}}=200\:\mathrm{GeV}$ $\mathrm{Au}+\mathrm{Au}$
collisions in PHENIX}
\authors{
{Jiamin Jin$^1$ (for the PHENIX Collaboration) %
\index{Jin, J.} % Abbreviated names of the author(s),
}\\[2.812mm]
{\normalsize \hspace*{-8pt}$^1$ Columbia University, New York, NY 10025, USA\\[0.2ex]
}}

\abstract{This poster presents the PHENIX results on high-$p_T$
$\gamma$-hadron and $\pi^{0}$-hadron correlations and shows the
modification of away side jet shape by the medium. By comparing
their jet shapes and yields, we see some evidence of the direct
$\gamma$ contribution to correlation functions.}


\makeindex
\begin{document}

\maketitle

\section{Introduction}\label{intro}

PHENIX has measured direct photon production in
$\mathrm{Au}+\mathrm{Au}$ at $\sqrt{s_{NN}}=200\:\mathrm{GeV}$
\cite{bib1}. Due to the suppression of high-$p_T$ pion production,
a large photon excess over meson decay background is seen in
central $\mathrm{Au}+\mathrm{Au}$ collisions at
$p_T>6\:\mathrm{GeV/c}$. The much higher statistics 2004 RHIC
$\mathrm{Au}+\mathrm{Au}$ run has been used to further explore the
possibility of constructing direct photon-hadron correlations.
Since direct photons, once produced, interact with the medium much
more weakly than hadrons, they provide a better measurement of the
energy and direction of the away side jets, thus we gain direct
information on the initial state effect and better control on the
modification of the away side jets by the strongly interacting
medium. The high $p_T$ photon sample PHENIX measures is a mixture
of direct photons and fragmentation (mostly $\pi^0$ decay)
photons. Therefore, it is important to study the high-$p_T$
$\gamma$-hadron and $\pi^{0}$-hadron correlations and compare them
systematically.

\section{High $p_T$ $\gamma$-hadron Correlations and $\gamma$ Per Trigger Yield}\label{techno}
The $\gamma$-hadron correlation analysis presented in QM2005
poster represents 1 billion minimum bias events. We use photons
with $p_T>5\:\mathrm{GeV/c}$ as leading particles and charged
hadrons with $1\mathrm{GeV/c}\:<p_T<5\:\mathrm{GeV/c}$ as
associated particles. The correlation function $C(\Delta\phi)$ is
constructed by pairing leading and associated particles.
$C(\Delta\phi)=\frac{N_{real}(\Delta\phi)}{N_{mix}(\Delta\phi)}$,
where $\Delta\phi$ is the difference of the azimuthal angles of
the pair. The $N_{real}(\Delta\phi)$ is built from pair members
belonging to the same event and $N_{mix}(\Delta\phi)$ is built
from pair members belonging to different events. This event mixing
technique removes the detector acceptance effects. The correlation
function can be decomposed into jets and flow components
\cite{bib2}:
\begin{equation}C(\Delta\phi)=J(\Delta\phi)+\xi(1+2v_{2}^{l}v_{2}^{a}\cos(2\Delta\phi))\end{equation}
where $J(\Delta\phi)$ is the jet contribution,
$v_{2}^{l}$($v_{2}^{a}$) is the elliptic flow for the
leading(associated) particles and $\xi$ is the background level.
The $v_2$ values are measured using reaction plane method, whereas
$\xi$ is fixed using ZYAM(Zero jet Yield At Minimum) assumption
\cite{bib2,bib3}. After removing the $v_{2}$ modulated background,
we correct the remaining $J(\Delta\phi)$ by single particle
efficiency and PHENIX $\Delta\eta$ acceptance, normalize it by the
number of triggers, thus obtain the photon per trigger yield.

\begin{figure*}[htb]
\vspace*{-7mm}
\begin{center}
\scalebox{0.27}
{\includegraphics{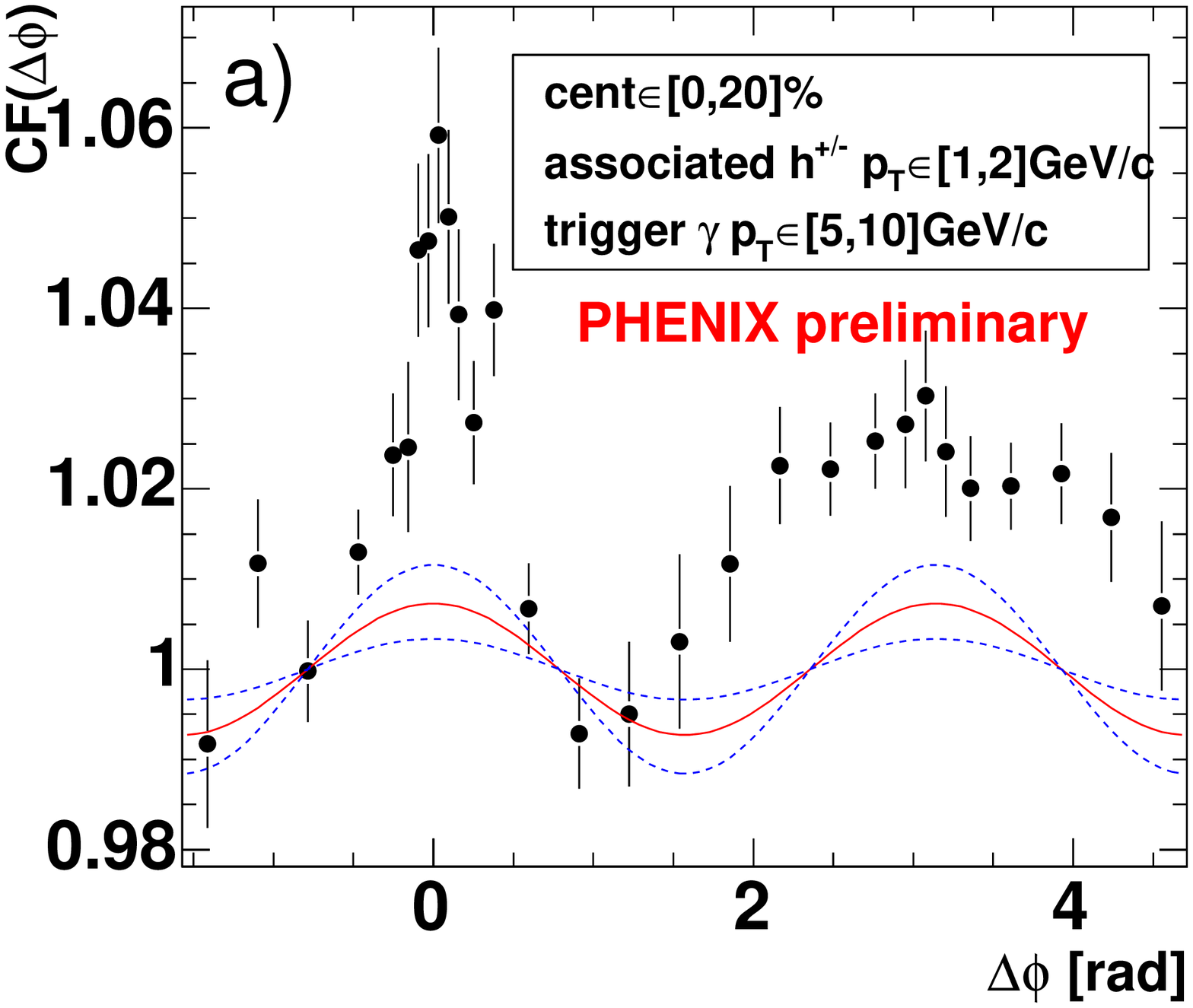}\includegraphics{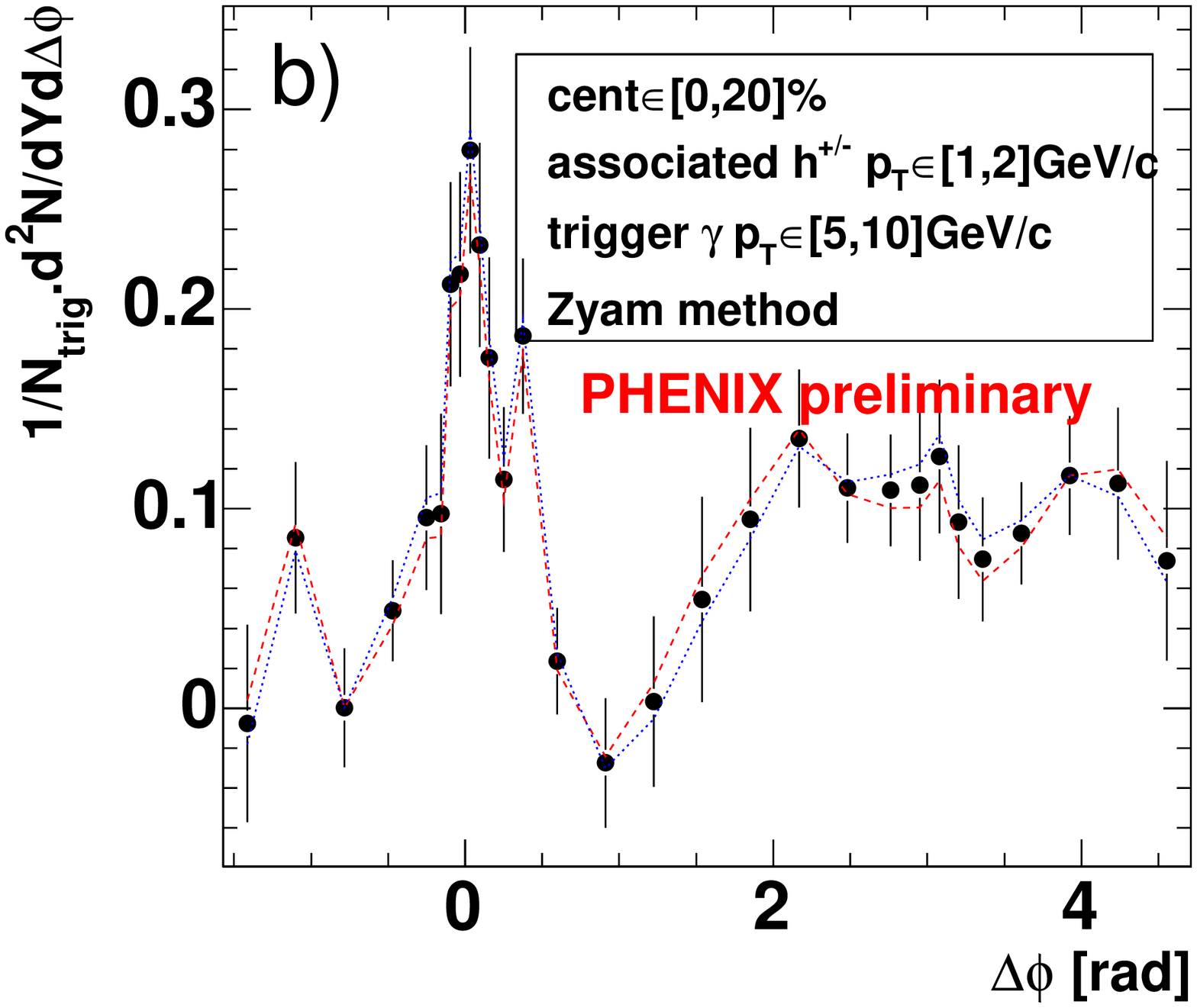}}
\vspace*{-7mm}
\begin{minipage}{12cm} \vspace*{-2mm} \caption{a) $\gamma$-hadron correlation function, lines
indicated the level of flow modulated background and its
systematic error band. b) Corresponding photon per trigger yield.}
\label{fig1}
\end{minipage}
\vspace*{-4mm}
\end{center}
\end{figure*}

Figure \ref{fig1} shows the typical $\gamma$-hadron correlation
function and per trigger yield for central
$\mathrm{Au}+\mathrm{Au}$ collisions. We see a strongly
modified(non-gaussian) away side shape.

\section{Results}\label{resu}
A simultaneous fit is applied to the correlation functions to
determine the jet and the flow contributions. The jet widths as
well as the near side jet yield are extracted directly from the
fit, whereas the away side yield is obtained by integrating the
correlation function over $\Delta\phi\in[\pi/2,3\pi/2]$, then
subtracting the background in the same range. This region is
chosen so that the overall flow contribution cancels out, thus
making the yield not sensitive to $v_2$.

\begin{figure*}[htb]
\vspace*{-1mm}
\begin{center}
\scalebox{0.4} {\includegraphics{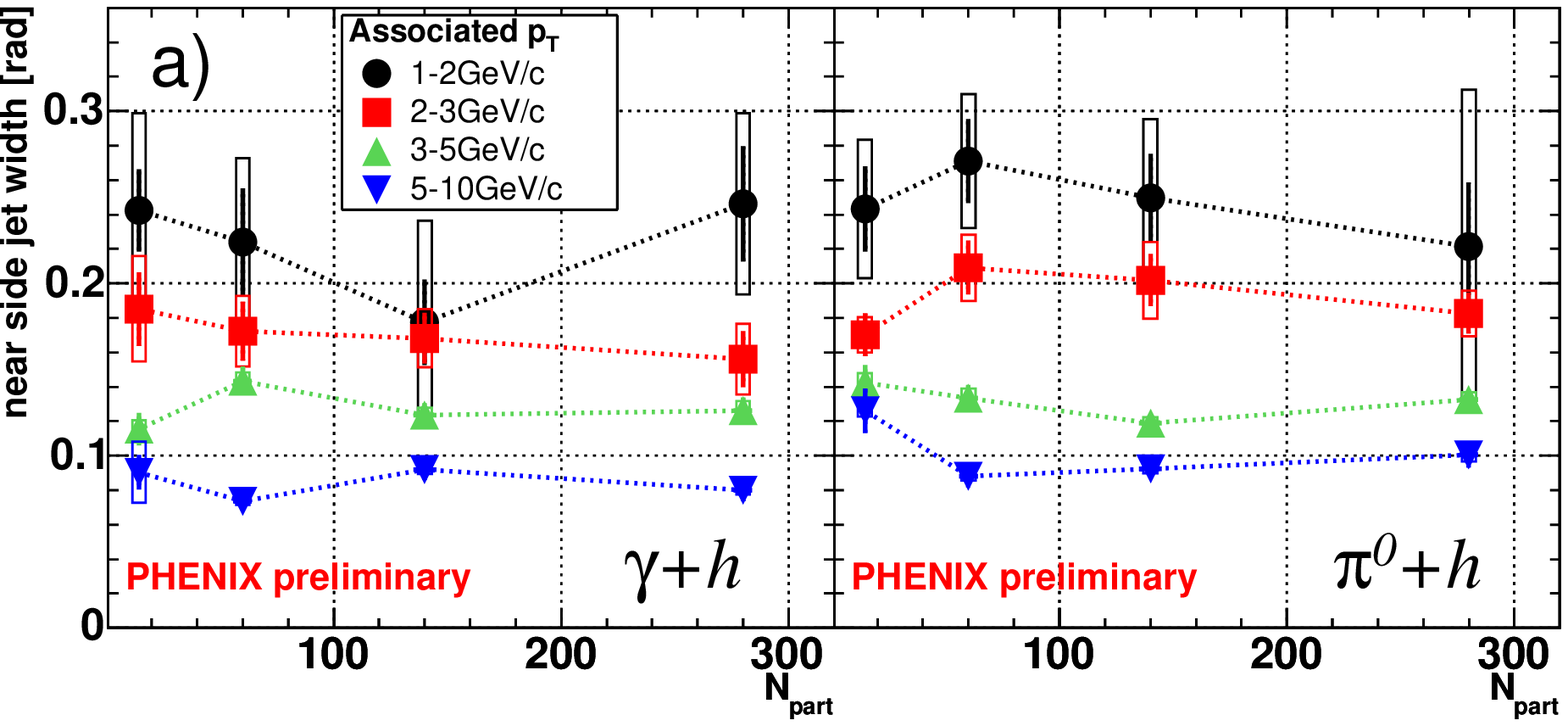}}
\scalebox{0.4}{\includegraphics{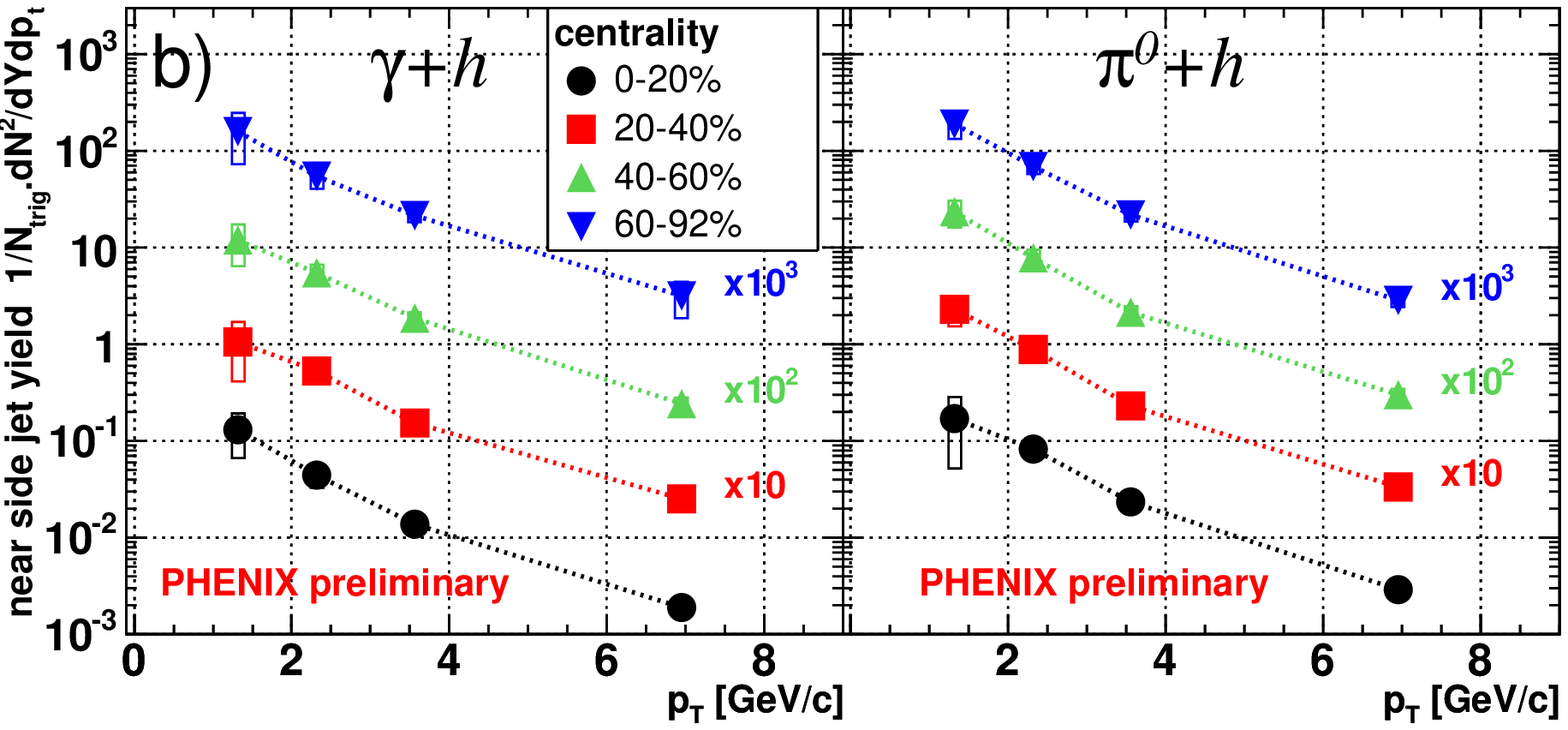}} \vspace*{-5mm}
\begin{minipage}{10cm} \vspace*{-1mm} \caption{a) comparison of near side width between $\gamma$ jet (left) and $\pi^{0}$ jet (right), b) same for near side yield} \label{fig2}
\end{minipage}
\vspace*{-3mm}
\end{center}
\end{figure*}

\begin{figure*}[htb]
\vspace*{-2mm}
\begin{center}
\scalebox{0.35} {\includegraphics{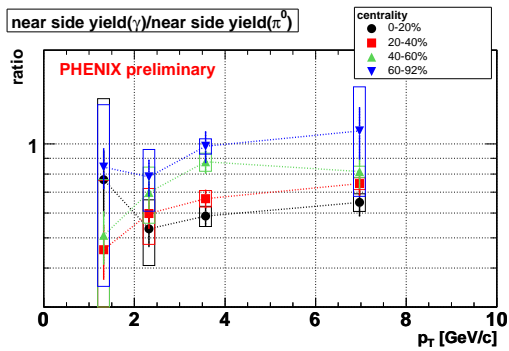}}

\begin{minipage}{13cm} \vspace*{-4mm} \caption{near side yield ratio between $\gamma$ jet and $\pi^{0}$ jet} \label{fig3}
\end{minipage}
\vspace*{-8mm}
\end{center}
\end{figure*}

PHENIX has measured $\pi^{0}$-hadron correlations in a similar way
with trigger $\pi^{0}$ $p_T>5\:\mathrm{GeV/c}$ \cite{bib4}. Figure
\ref{fig2} shows the comparison of near side jet width and yield
between $\gamma$-hadron and $\pi^{0}$-hadron correlations. They
have comparable widths, whereas $\gamma$ jet yields are
consistently lower than $\pi^{0}$ jet yields. Figure \ref{fig3}
plots the ratio of $\gamma$ near side jet yield over $\pi^{0}$
near side jet yield as a function of associated particle $p_T$ for
different centrality bins. The ratios are all below one. Direct
photons contribute to this yield reduction since they are not from
jets, thus having zero yield on the near side. The ratios also
become smaller for more central bins. This trend is consistent
with the results in \cite{bib1}, which established that there is
larger direct photon signal in central collisions.

\begin{figure*}[htb]
\vspace*{-1mm}
\begin{center}
%\begin{minipage}{15cm}
%\scalebox{0.22}
%{\includegraphics{bothrun4_10_0_10.eps}\includegraphics{bothrun4_11_0_10.eps}\includegraphics{bothrun4_9_0_12.eps}}
%\end{minipage}
\scalebox{0.46}{\includegraphics{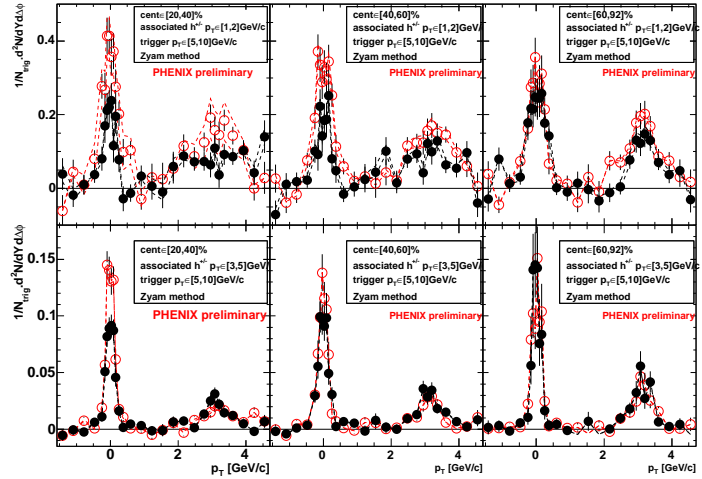}} \vspace*{-5mm}
\begin{minipage}{13cm} \caption{Upper row: $\gamma$-hadron (filled points) and $\pi^{0}$-hadron (open circles) correlations for 3 different centrality bins and $1$--$2\:\mathrm{GeV/c}$ associated $p_T$ bin. Lower row: same for $3$--$5\:\mathrm{GeV/c}$ associated $p_T$ bin. } \label{fig4}
\end{minipage}
\vspace*{-3mm}
\end{center}
\end{figure*}

Figure \ref{fig4} shows $\gamma$-hadron and $\pi^0$-hadron
correlation functions on top of each other for 3 centrality bins
and 2 associated particle $p_T$ bins. It allows us to do a direct
comparison and have some feeling about the subtle jet shape change
between the two. In addition to the near side jet yield reduction
noted above, the upper panels show that the away side peaks in
$\gamma$-hadron correlations seem to have a stronger distortion at
low $p_T$, whereas the lower panels show that once the $p_T$ of
the associated hadrons reach 3 $\mathrm{GeV/c}$, the two
correlations appear similar on the away side.

\section{Conclusions}\label{concl}
High $p_T$ $\gamma$-hadron correlation functions are measured in
different centrality bins and associated particle $p_T$ bins. A
distortion on the away side jet due to the coupling of the hadrons
and the strong interacting medium is observed. Comparison between
$\gamma$-hadron and $\pi^{0}$-hadron correlations shows a decrease
of the near side $\gamma$ jet yield and a stronger modification on
the away side $\gamma$ jet, both effects can be attributed to the
presence of direct photons. Further investigation on the role
direct photon plays in correlation studies is undergoing.

\vfill\eject
\end{document}